\documentclass[11pt]{article}
\usepackage{moriond,epsfig}

\def\sun{\odot}
\def\la{\, \lower2truept\hbox{${< \atop\hbox{\raise4truept\hbox{$\sim$}}}$}\,}
\def\ga{\, \lower2truept\hbox{${> \atop\hbox{\raise4truept\hbox{$\sim$}}}$}\,}

\def\be{\begin{equation}}
\def\ee{\end{equation}}
\def\bea{\begin{eqnarray}}
\def\eea{\end{eqnarray}}

\begin{document}
\vspace*{4cm}
\title{Bright Side versus Dark Side of Star Formation
       --- UV and IR Views}

\author{C.K. Xu and V. Buat}

\address{IPAC, Caltech, 770 S. Welson Ave., Pasadena, CA91125, USA. \\
Laboratoire d'Astrophysique de Marseille, BP 8, Traverse
du Siphon, 13376 Marseille Cedex 12, France.}



\maketitle\abstracts{This is a review talk on the UV and infrared
selected galaxies. The central question
addressed is: do UV and infrared surveys
see the 2 sides of star formation of the same population, or star
formation of 2 different populations? 
We first review the literature
on the UV and IR selected galaxy samples, try to quantify the
difference and overlaps between these two populations of star forming 
galaxies. We then present some preliminary results of
a GALEX/SWIRE comparison study for IR and UV selected galaxies
at z=0.6, in an attempt to constrain the evolution of
the dust attenuation and of stellar mass of these galaxies.}

\section{Introduction}
The evolution of star forming galaxies tells much about the history of
the universe. The star formation activity in these galaxies can be
best studied by observing the emission from young massive stars in the
rest frame UV and FIR. The UV observations record the direct light
from the hot young stars, and the FIR observations collect star light
absorbed and then re-emitted by the ubiquitous dust. A complete
picture of star formation in the universe can only be obtained when
the observations in these two wavebands are properly synthesized.
Indeed, our knowledge on the star formation history of the universe
('Madau diagram') has been mostly derived from deep surveys in the
rest frame UV and FIR. Many studies have been devoted to methods of
deriving star formation rate of individual galaxies using the UV or FIR 
luminosities, particularly concerning the correction for the dust 
attenuation~\cite{xu95}~\cite{buat96}~\cite{wang96}~\cite{calzetti97}~\cite{meurer99}~\cite{jorge05a}.
The strengths and shortcomings of these methods have been discussed
thoroughly in the 
literature~\cite{kennicutt98}~\cite{adelberger00}~\cite{bell02}~\cite{bell03a}~\cite{buat05}~\cite{kong04}~\cite{jorge05a}.  
However, an arguably more important issue is the
selection effect of the surveys that can be summed up by the following
question: Do UV and IR surveys see the two sides ('dark' and 'bright')
of the star formation of the same population of galaxies, or they see
two different populations of star forming galaxies?  This is important
because if the correct answer is the latter, then even
if one can estimate accurately the star formation rate for galaxies in
surveys in one band, the star formation in galaxies detected
in the other band is still missing. 
Actually this question is in the core of an
on-going debate on whether the SFR of z$\sim 3$ universe can be
derived from observations of Lyman-break galaxies (LBGs), which are UV
selected star forming galaxies at z$\sim 3$~\cite{adelberger00},
given that SCUBA surveys in sub-millimeter (rest frame FIR for
z$\ga 2$) detected many violent star forming galaxies at about the
same redshift that are not seen by LBG surveys~\cite{smail01}~\cite{smail04}.

This talk is arranged as follows: we'll first concentrate on local
UV and IR selected galaxies and exam how much these two samples
overlap. We compare their infrared to UV ratios, total luminosities,
Hubble types, stellar mass, and the clustering behaviour. The
comparison between the UV luminosity function for the IR selected galaxies
and the GALEX UV luminosity function
tells whether IR surveys miss a substantial population of UV galaxies.
Similarly the infrared luminosity function for UV selected
galaxies, when compared to IRAS luminosity function,
tells what kind of IR galaxies are missed by the UV
surveys. We'll exam in detail a population of IR-quiet UV galaxies, and
compare luminous UV galaxies, the so called UVLGs that are recently
discovered by GALEX, with ULIRGs. I'll then give a brief review on the
literature of LBGs and SCUBA galaxies, focusing on
their comparisons as UV and IR selected galaxies at redshift about
3. The last major topic is on UV and infrared galaxies at intermediate
redshifts. Here I'll report early results of a GALEX/SWIRE comparison
study on galaxies in the redshift range of 0.5 to 0.7, selected using
photometric redshifts. We'll investigate evidence for evolution of the
extinction in UV and IR selected galaxies, and for evolution of
stellar mass in these galaxies.  Then I'll wrap it up with a
summary.

\section{Local UV and IR Galaxies: How Much Do They Overlap?}

Martin et al.~\cite{martin05} derived the infrared-UV bivariate luminosity
function of local UV plus IR galaxies. There appears to be a
saturation of UV luminosity at about 2 $10^{10}$ L$_\sun$,
beyond which the density drops fast. The distribution seems to be
bi-modal, with galaxies of relatively low IR/UV ratios are reasonably
well separated from those with high ratios.  
There is a strong dependence of the IR-to-UV ratio (the
best indicator of the UV attenuation) on L$_{tot}$, which is the sum
of the UV luminosity and infrared luminosity.  Actually, the ratio
increases almost proportionally with L$_{tot}$.  The L$_{tot}$ luminosity
function of local UV plus IR galaxies has the form of log-normal.  It
appears that UV galaxies are absent in the high L$_{tot}$ end 
($>$ a few $10^{11}$ L$_\sun$).

Buat et al.~\cite{buat05} compared the dust attenuation properties of
NUV and FIR selected samples selected from GALEX and IRAS databases, 
respectively. The median value of the attenuation
in NUV is found to be ~1 mag for the NUV-selected sample, versus ~2 mag for
the FIR-selected one. Within both samples, the dust attenuation is found to
correlate with the luminosity of the galaxies. These results are 
consistent with the pre-GALEX study of Iglesias-P\'{a}ramo et al~\cite{jorge04}
using UV data of FOCA observations.

In order to exam the difference and overlaps between UV and IR
selected galaxies quantitatively, we carried out detailed analysis in two local
samples, one is Infrared selected and the other UV selected. These are
the same samples discussed in Jorge Iglesias-P\'{a}ramo's talk~\cite{jorge05b}, so I
refer you to that paper for the details of the samples. Here I just
mention the major selection criteria. The IR selected sample is taken
from the PSCZ catalog, which are IRAS galaxies brighter than 
$f_{60} = 0.6$
Jy and they all have measured redshifts. The UV sample is selected
from galaxies brighter than NUV=17 magnitude found in the fields
observed in the G1 stage of GALEX mission.  In Fig.1 we show the
IR-to-UV luminosity ratio distributions of UV and IR galaxies. They
look rather different from each other. The overlap between the two
distributions is only about 30\%. The mean ratio of IR galaxies is one
order of magnitude higher than that of the UV galaxies. The IR-to-UV
ratio can be directly translated to the UV attenuation.  As reported
by Buat et al.~\cite{buat05}, the average UV attenuation of UV galaxies is
only about 1 magnitude, while for IR selected galaxies, the average UV
attenuation is above 2 magnitude.

\begin{figure}
\centerline{\psfig{figure=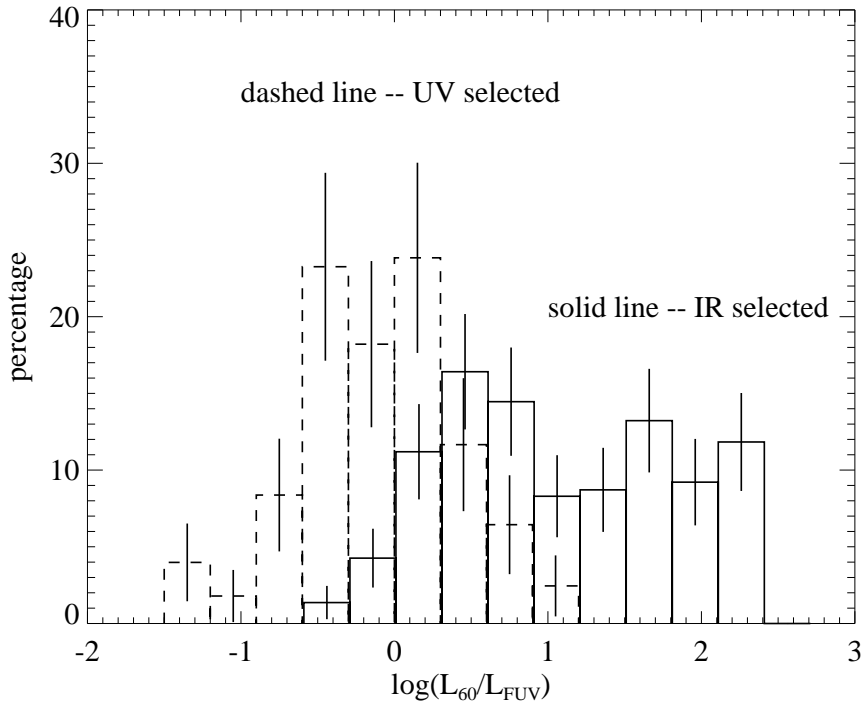,height=4in}}
\caption{FIR/UV distributions of UV and IR galaxies.
\label{fig:fig1}}
\end{figure}

 Fig.2 shows the IR/UV ratio versus
the L$_{tot}$ for individual UV and IR galaxies. In this plot, the two
populations are also separated, with the IR galaxies taking the high
L$_{tot}$, high IR-to-UV ratio end of the correlation, and the UV
galaxies occupying mostly the lower end of the correlation.  This
result can be explained by the strong correlation between
the ratio and the L$_{tot}$, and
the selection effect which biases the IR
and UV samples toward the high and low ends of IR/UV ratios,
respectively.

\begin{figure}
\centerline{\psfig{figure=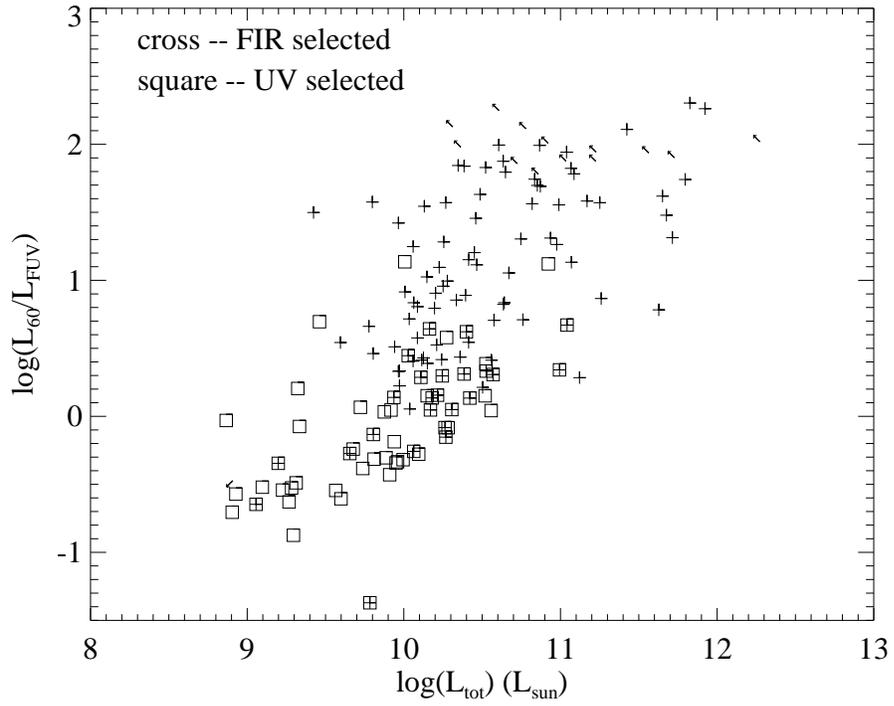,height=4in}}
\caption{FIR/UV vs. L$_{tot}$ of individual galaxies.
\label{fig:fig2}}
\end{figure}

In Fig.3 we compare the Hubble type distributions
of UV and IR galaxies. The overlap between the two
distributions is $\sim 60\%$. There is a
significant excess of Pec/Int/merg galaxies in the FIR selected
sample (39\%) compared to those in the UV selected sample (14\%). For
normal galaxies both UV and FIR selected samples peak in the bin of
Sab/Sb/Sbc. Detailed analysis shows that the median type for normal UV
galaxies is Sc and that of normal FIR galaxies is Sb. 

\begin{figure}
\centerline{\psfig{figure=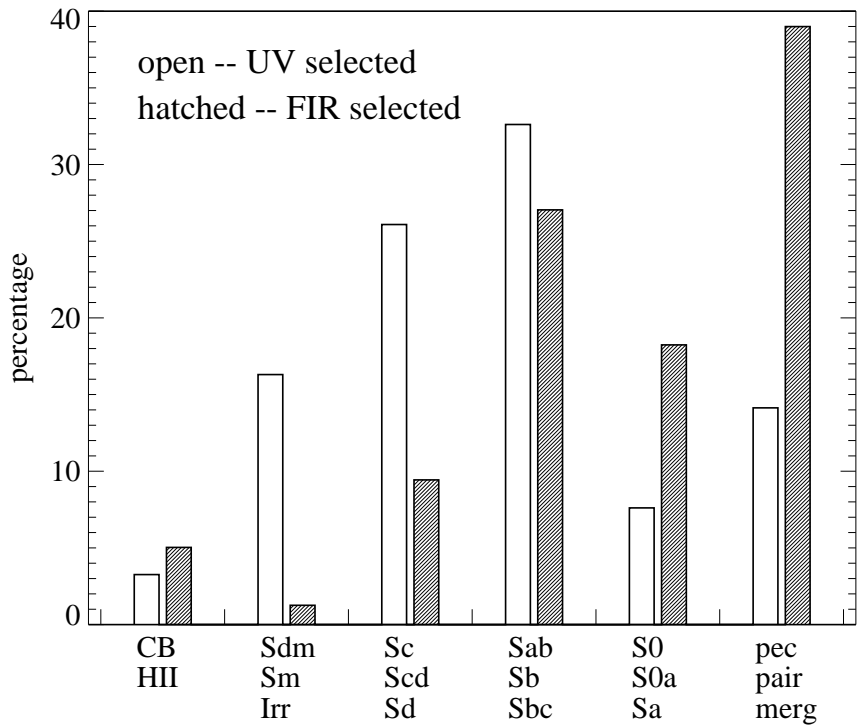,height=4in}}
\caption{Morphological distributions of UV and IR galaxies.
\label{fig:fig3}}
\end{figure}

Heinis et al.~\cite{heinis04} carried out the angular correlation 
analysis for the UV population,
using FOCA data. They found a correlation length of 
3.2 (+0.8, -2.3) Mpc (H$_0$/100)$^{-1}$. 
Compared to the correlation length
of IRAS galaxies determined by Strauss et al.~\cite{strauss92}, 
which is 3.9$\pm 1.8$ Mpc (H$_0$/100)$^{-1}$, 
the UV galaxies seem to be slightly less clustered
than IR galaxies, consistent with the fact that UV galaxies are
preferentially later type spirals and irregulars.

\begin{figure}
\centerline{\psfig{figure=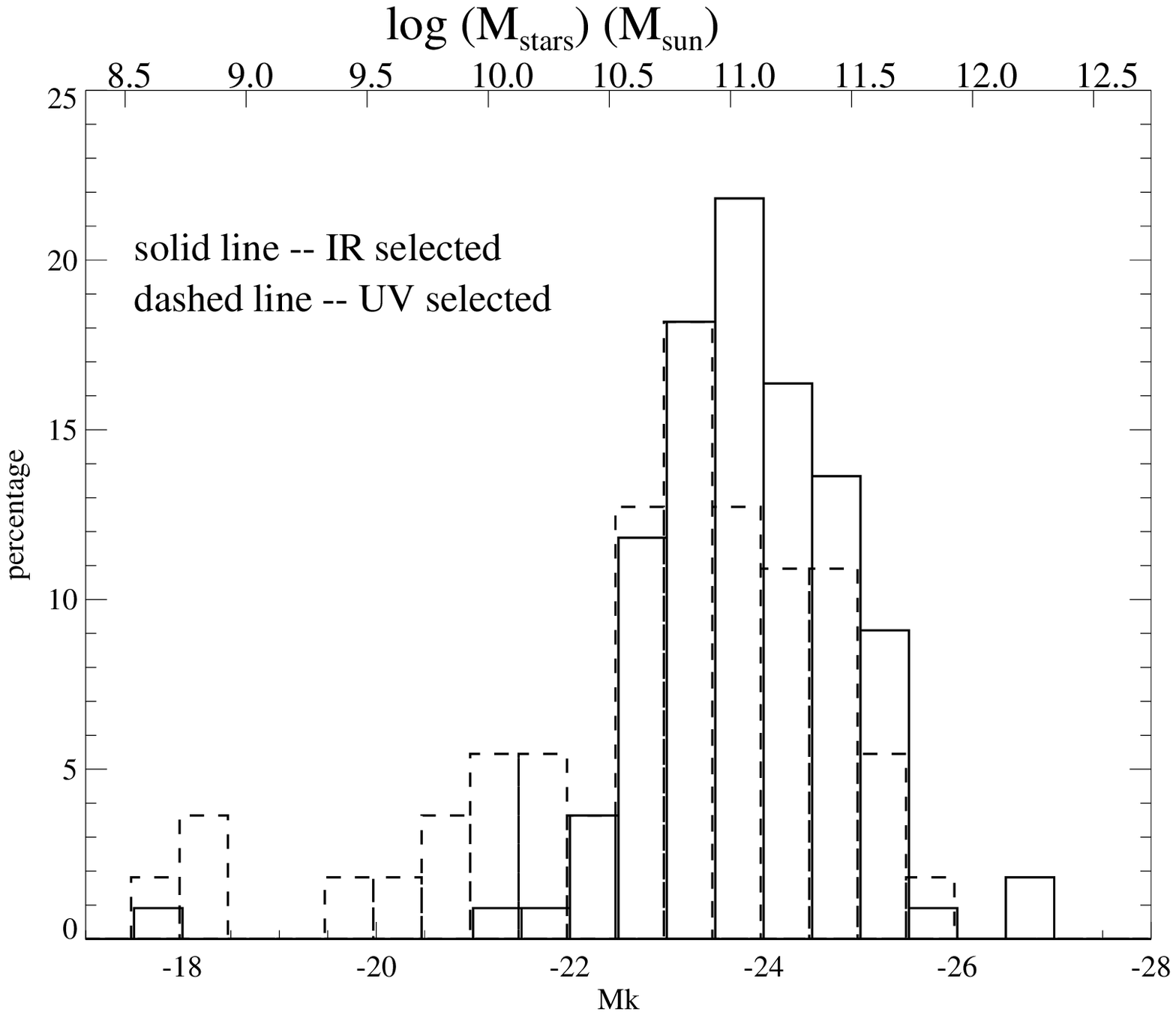,height=4in}}
\caption{Stellar mass distributions of UV and FIR selected samples.
\label{fig:fig4}}
\end{figure}

Recent literature on galaxy formation and evolution has revealed the
stellar mass as a fundamental variable in the characterization of
galaxy populations~\cite{boselli01}~\cite{kauffmann03}.  And
the K-band luminosity is the best estimator of the stellar 
mass~\cite{bell01}~\cite{bell03b}. Most of the galaxies in our UV
and IR selected samples have been detected in K band by 2MASS.  For a
few sources that are not detected by 2MASS, we used the so called
survival technique\cite{kaplan58}~\cite{feigelson85}~\cite{schmitt85}
to exploit the information content in the upper
limits. The results are plotted in Fig.4. The solid histogram is the
stellar mass distribution of IR galaxies and the dashed histogram is
that of UV galaxies.  The median K band
absolute magnitude of the UV selected is -21.23 and that of the IR
selected sample -21.95. The conversion factor $\rm M_{stars}/L_K =
1.32 M_\sun/L_\sun$, which is derived for a stellar population 
with constant star formation rate and a Salpeter IMF~\cite{cole01},
is assumed when converting the
K band luminosity to stellar mass.  The median K band absolute
magnitudes correspond to median stellar mass of $10^{10.75}$ M$_\sun$
for the UV selected sample and of $10^{11.04}$ M$_\sun$ for the FIR
selected sample, respectively.  Both medians are slightly lower than
the mass corresponding to the K band L$_*$ of 2MASS galaxies~\cite{cole01}.
Here again there is a good overlap between the two
populations, the medians differ only by less than a factor of 2.
On the other hand, IR galaxies are slightly tilted
for the more massive end, and more UV galaxies have relatively low
mass.

The two plots in Fig.5 show that (1) UV galaxies which have the lowest
mass also have the lowest L$_{tot}$ and the lowest IR-to-UV ratio; (2)
most massive IR galaxies are not the galaxies with the highest
L$_{tot}$; (3) the brightest galaxies have mass about M$_*$; (4) for
given mass, UV galaxies have lower L$_{tot}$ and IR-to-UV ratio than IR
galaxies.

\begin{figure}
\vbox {
  \begin{minipage}[l]{1.0\textwidth}
   \hbox{
      \begin{minipage}[l]{0.5\textwidth}
       {\centering \leavevmode \epsfxsize=\textwidth 
        \epsfbox{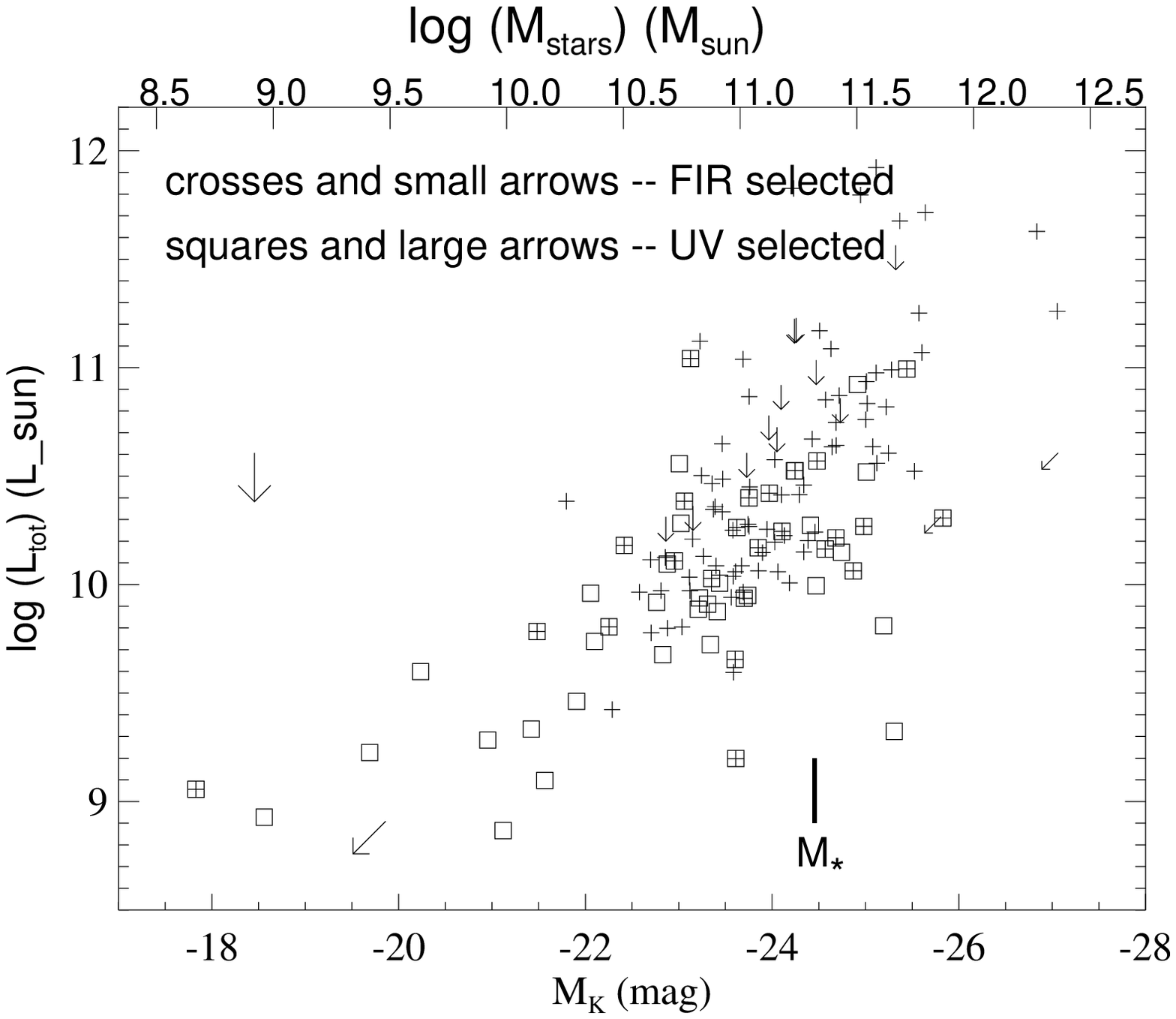}}
      \end{minipage} \  \hfill \
      \begin{minipage}[r]{0.5\textwidth}
       {\centering \leavevmode \epsfxsize=\textwidth 
        \epsfbox{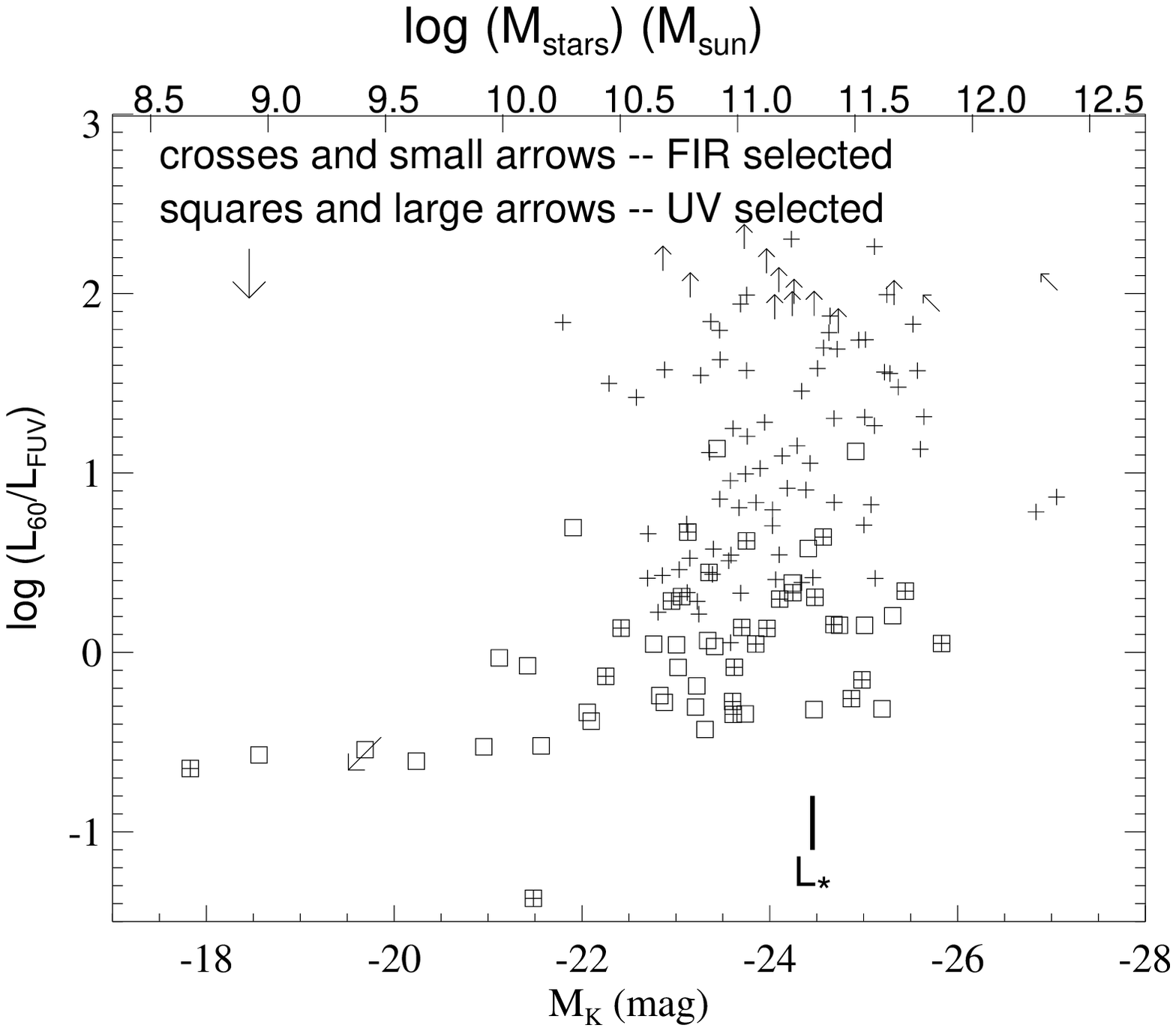}}
      \end{minipage} \  \hfill \
   }
  \end{minipage} \  \hfill \
\begin{minipage}[l]{1.0\textwidth}
\caption{L$_{tot}$ vs. M$_{star}$ and
FIR/UV vs. M$_{star}$ of UV and IR galaxies.}
\end{minipage}
}
\label{fig5}
\end{figure}

In order to check how much UV and IR galaxies overlap
and how much they miss each other, we have derived infrared luminosity
function of UV galaxies, as plotted in Fig.6a by the open squares with error
bars. It is compared it with the IRAS luminosity function shown by the
solid line. The UV luminosity function of IR selected galaxies, as
plotted in Fig.6b by the open diamonds with error bards, is compared
to the GALEX luminosity function shown by the solid line. 
It appears that in the UV selected sample, galaxies of
infrared luminosity larger than $10^{11}$ solar luminosity are 
 substantially under-represent (the
ULIRGs being completely absent). In contrast, all UV
galaxies brighter than $10^9$ solar luminosity are fully represented in
the IR selected sample, although some fainter UV galaxies could be missing
in the IR sample.

\begin{figure}
\vbox {
  \begin{minipage}[l]{1.0\textwidth}
   \hbox{
      \begin{minipage}[l]{0.5\textwidth}
       {\centering \leavevmode \epsfxsize=\textwidth 
        \epsfbox{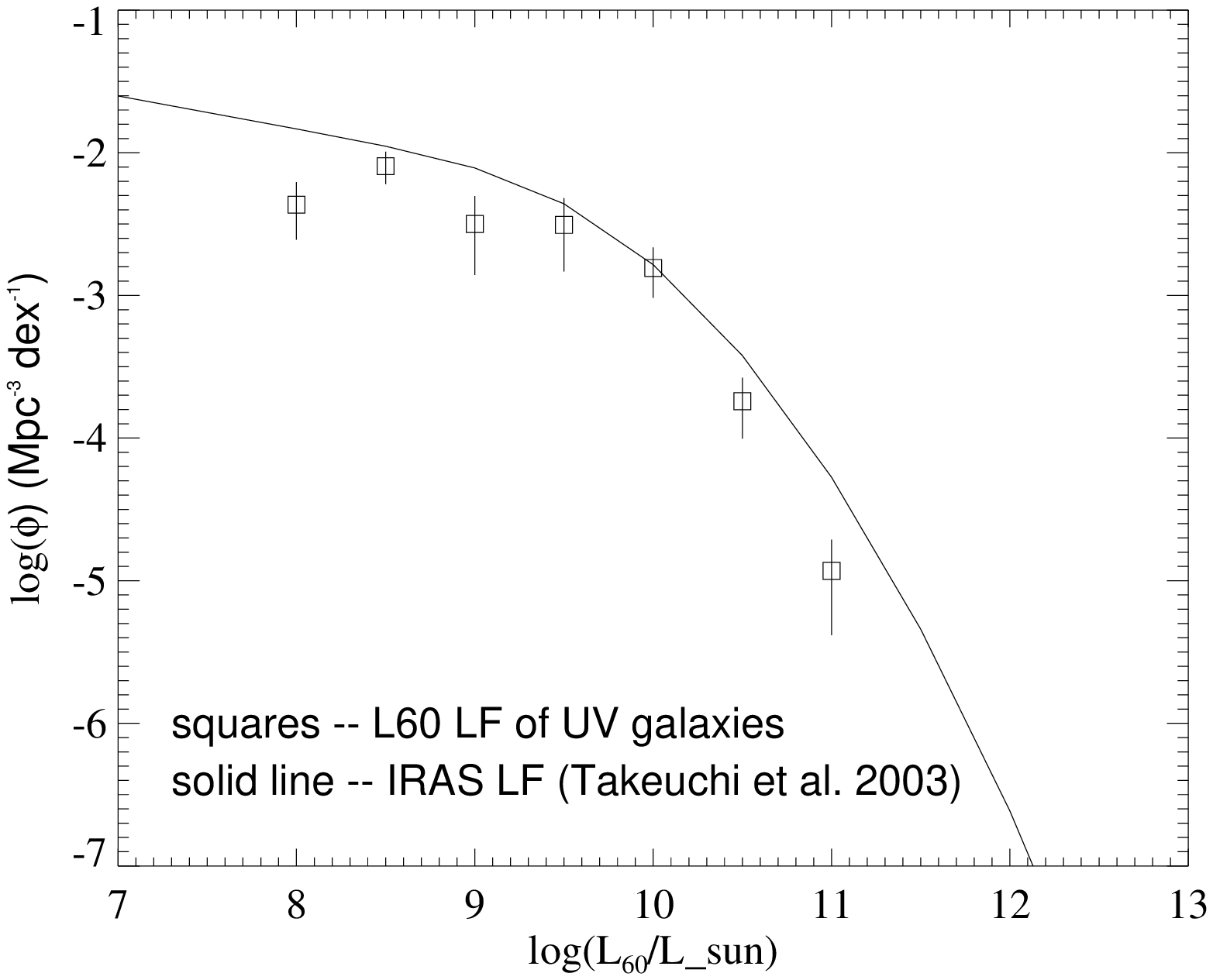}}
      \end{minipage} \  \hfill \
      \begin{minipage}[r]{0.5\textwidth}
       {\centering \leavevmode \epsfxsize=\textwidth 
        \epsfbox{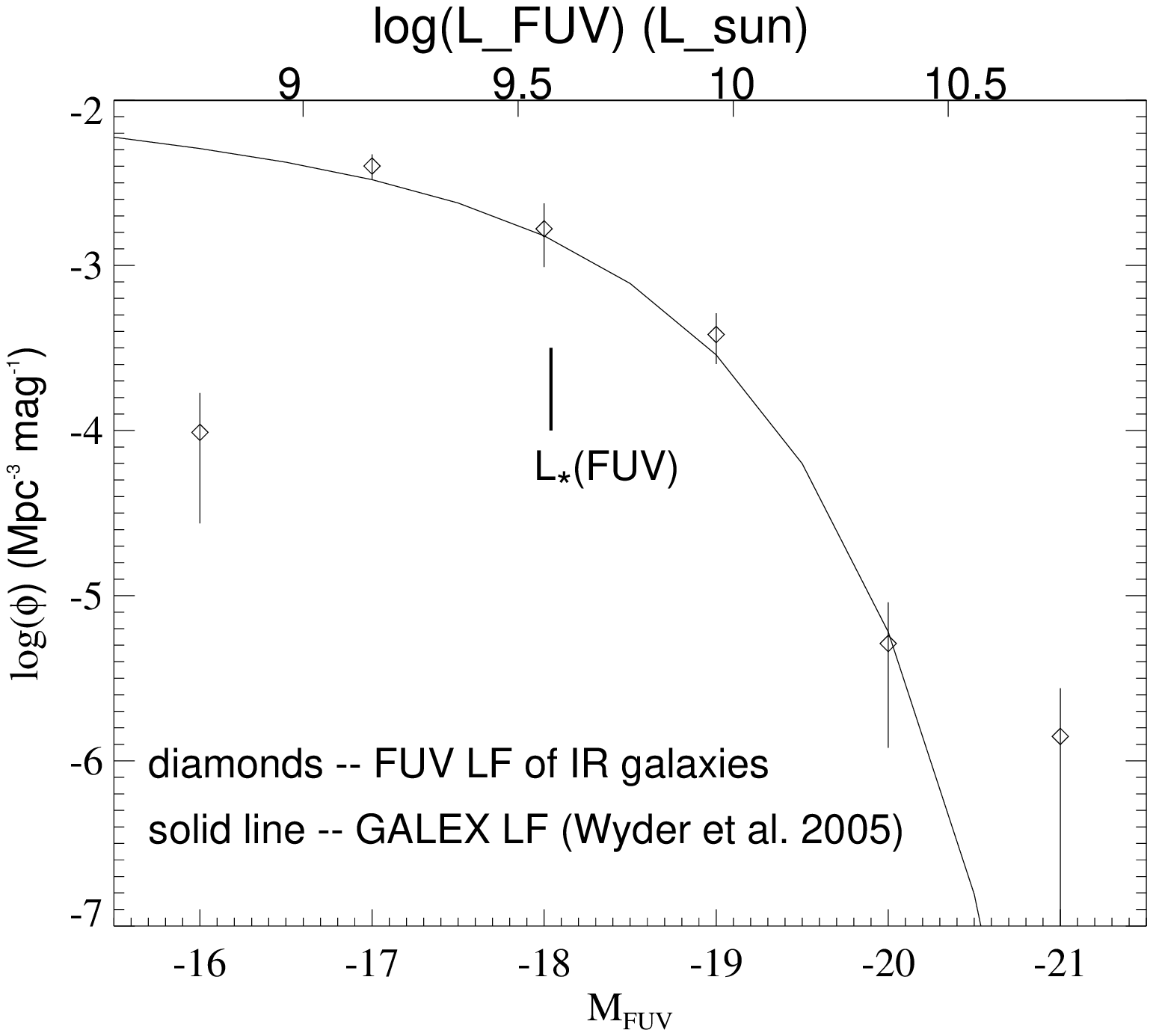}}
      \end{minipage} \  \hfill \
   }
  \end{minipage} \  \hfill \
\begin{minipage}[l]{1.0\textwidth}
\caption{UV and IR luminosity functions. The GALEX FUV (1530{\AA})
luminosity function is taken from Wyder et al. (2005), and the
IRAS 60$\mu m$ luminosity function is taken from Takeuchi et 
al. (2003).}
\end{minipage}
}
\label{fig6}
\end{figure}

\section{Special Populations of UV Galaxies}
The UV galaxies missed by IR surveys are so called `IR-quiet'
star-forming galaxies. The proto-type 
is the famous low metallicity dwarf I-Zw-18.  It has the lowest
metallicity (1/50th of solar) known for galaxies,
and its baryonic mass is only about 2 $10^8$ M$_\sun$.
Its FUV luminosity as measured by GALEX is 2.5 $10^8$ L$_\sun$. 
I-Zw-18 has never been
detected in far-infrared. The IRAS upperlimit corresponds to an
upperlimit for the IR-to-UV ratio of less than 0.25.

Another prototype IR-quiet galaxy is SBS-0335-052. 
It has the second lowest known
metallicity of 1/35th solar. The mass is higher than I Zw 18, about 
2 $10^9$ M$_\sun$. It was undetected by IRAS, but detected by both
ISO and Sptizer. Its IR SED reported by 
Houck et al.\cite{houck04} is
very different from that of ordinary galaxies:  its $f_{60\mu m}$ 
is about the same as f$_{25\mu m}$ whereas normal galaxies such as
the Milky Way has the $f_{60\mu m}/f_{25\mu m}$ ratio $\sim$5 -- 10. 
In summary, IR quiet galaxies are dwarf galaxies of low
metallicity, usually lower than 1/10th solar. They have relatively low
mass and low UV luminosity. They are about 10 times fainter than the 
L$_*$ of FUV, and about 100 times fainter than the Lyman Break
Galaxies (LBG). Therefore they are no local counterparts of LBGs.
Typically they have $\rm L_{60}/L_{UV} < 0.3$.
And they are less than 15\% of galaxies in a UV selected sample. 

The true local counterparts
of Lyman break galaxies are a population of UV luminous galaxies, or
UVLGs, recently discovered by GALEX~\cite{heckman05}.
These galaxies are brighter than L$_{FUV}=2 \times 10^{10}$ L$_\sun$. 
And they have a density about 100 times lower than that of LBGs, at  
$\sim 10^{-5}$ Mpc$^{-3}$. These galaxies can be
divided into compact galaxies having higher surface brightness and
lower mass, and the large galaxies having lower surface brightness and
larger mass. The compact galaxies
are similar to LBGs in terms of size and mass.  Also, compact galaxies
and LBGs have similar UV attenuation, star formation history parameter
b, and the metallicity, while large galaxies have values in these
variables differ from that of LBGs.  Heckman et
al.~\cite{heckman05} identified compact 
UVLGs as the local counterparts of Lyman
Breaks. In Fig.7, the UVLGs are compared with other
population of UV and IR galaxies. By definition, UVLGs
occupy the bright end of the UV population.
But still, their L$_{tot}$ does not go much beyond
$10^{11}$ L$_\sun$, never being as  bright as ULIRGs.
It appears that UVLGs have the dust
attenuation in the same range as that of main
population of UV selected galaxies. However, it should
be pointed out that this result can be rather
uncertain because, due to the lack of FIR data,
the dust attenuation of UVLGs is estimated through SED fitting. 

\begin{figure}
\centerline{\psfig{figure=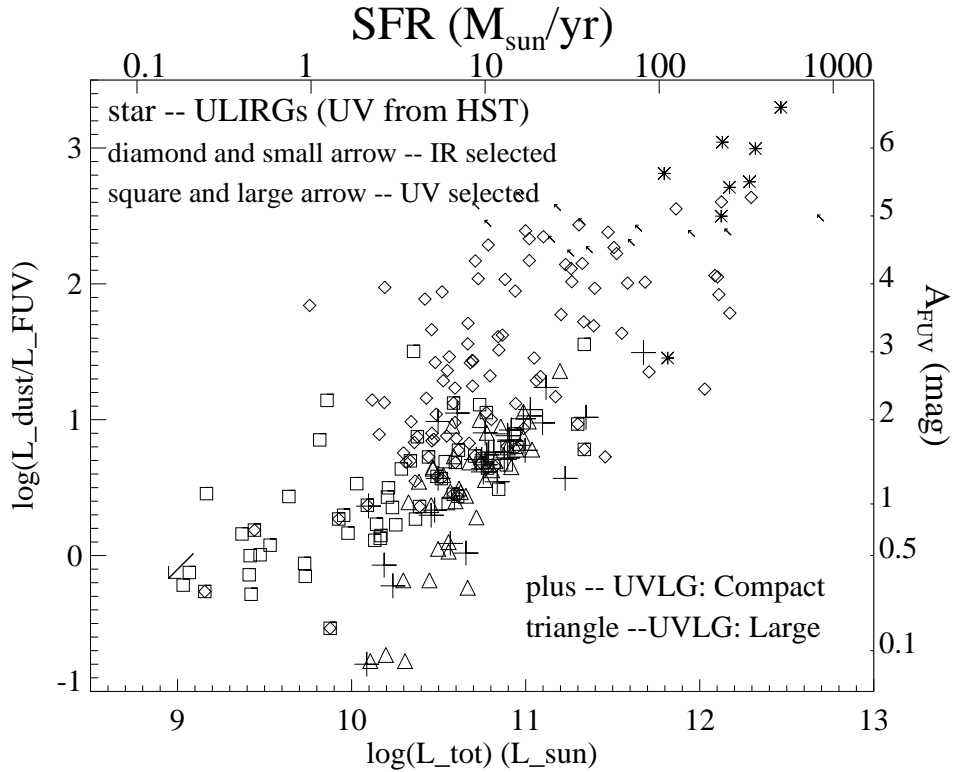,height=4in}}
\caption{Comparison of UVLGs with other UV and IR
populations. The data for UVLGs are taken from
Heckman et al. (2005), and the data for
ULIRGs are taken from Trentham et al. (1999)
and Goldader et al. (2002).
\label{fig:fig7}}
\end{figure}

Lyman Break Galaxies (LBGs) and SCUBA
galaxies are UVLGs and ULIRGs at z$\sim 3$. 
There is little overlap between these two populations.
The SCUBA survey of LBGs by Chapman et
al.~\cite{chapman00} has only 1 detection. The works by Adelberger \&
Steidal~\cite{adelberger00} and Chapman et al.~\cite{chapman04} show that
nearly all LBGs have IR/UV ratio less than 100 while
nearly all SCUBA galaxies have the ratio larger than 100. 
Compared to LBGs, SCUBA galaxies are 
heavier~\cite{smail04}
and more strongly clustered~\cite{blain04}.

\section{UV and IR Galaxies at $z= 0.6$: GALEX/SWIRE Comparison}

The first question is: why z=0.6? There are several reasons
for us to concentrate on this redshift. First of all, z=0.6 is close
to the peak of cosmic star formation suggested by the
SDSS fossil studies of local galaxies~\cite{heavens04}. 
Secondly, for larger redshift,
the NUV band of GALEX is affected by the rest frame Ly$_\alpha$ emission
or absorption, therefore the K-corrections can be very uncertain. And
finally, at z=0.6, there are several coincidences, which make the K
corrections very straightforward: (1) the GALEX NUV band coincides
with the rest frame FUV; (2) the MIPS 24$\mu m$ band measures
the rest frame 15$\mu m$ emission, which is an infrared luminosity
indicator extensively studied by ISO. And finally, the IRAC 3.6$\mu m$
band flux measures the rest frame K-band emission which is the best
stellar mass indicator. 

The field studied is the GALEX ELAIS-N1\_00 which is inside the SWIRE
ELAIS N1 field, covering 0.6 deg$^2$ of sky.  The nominal 5$\sigma$
sensitivity limit of SWIRE 24$\mu m$ survey is $f_{24}=0.15$ mJy, but
below $f_{24}=0.2$ mJy the catalog becomes progressively 
incomplete~\cite{surace04}~\cite{shupe05}. The GALEX NUV data are
confusion limited at NUV$\simeq 24$. The photometric redshifts,
derived using optical UgriZ
magnitudes obtained in the ELAIS-N1 optical survey and
SWIRE 3.6 -- 24$\mu m$ flux densities,
are taken from Rowan-Robinson et al.~\cite{mrr05} which have rms
deviation of $\log_{10}(1+z)$ about 10\%.
In the area considered here, 
we select a sample of 1124 NUV sources which
are brighter than NUV=24 and which have the photometric redshifts
in the range of $0.5 \leq z \leq 0.7$, and another sample of
316 24$\mu m$ sources brighter than $f_{24}=0.2$ mJy in
the same photometric redshift range. Among the $z\sim 0.6$ 
24$\mu m$ sources, 127(40\%) are detected by GALEX in NUV, and
the 24$\mu m$ detection rate of the z$\sim 0.6$ NUV sources
is only 14\%. For GALEX sources not detected in 24$\mu m$ 
band, an upperlimit of $f_{24}=0.2$ mJy is assigned. 
For SWIRE sources not detected by GALEX, the NUV upperlimit
corresponds to NUV=24 mag. All galaxies in both samples are detected
by SWIRE in the 3.6$\mu m$ band with the nominal sensitivity
limit (5$\sigma$) of $f_{3.6}=3.7\; \mu$Jy.
The FUV luminosities ($\nu L_\nu (1530{\AA})$) are derived from
the NUV magnitudes and the photometric redshifts (hereafter photo-z). 
Given the large errors in photo-z, the k-correction related 
uncertainties are neglected. In the same manner, 
we assume the Spitzer 24$\mu m$ observations measure the
rest frame $15\mu m$ emission in these galaxies, and the
total dust luminosity is estimated using the conversion
factor $L_{dust}= 11.1\times L_{15}$~\cite{chary01}.
The rest frame K band (2.2$\mu m$) luminosities of these
galaxies are calculated using the Spitzer 3.6$\mu m$ flux
densities and the photo-z. The stellar mass is estimated
from the K-band luminosity 
using the mass-to-light ratio $\rm M_{stars}/L_K =
1.32 M_\sun/L_\sun$~\cite{cole01}.

Because the $f_{24}$
detection rate of the z=0.6 UV sources is only 14\%, the only way we
can get meaningful information about the IR emission of these UV
galaxies is through stacking. We binned the UV galaxies into these 4
luminosity bins. Stacking the images of galaxies in each bins,
we derived mean $f_{24}$ and the mean L$_{dust}$/L$_{FUV}$.
The latter are compared with the values of z=0 galaxies
in Fig.8a. There is no significant difference between the
mean L$_{dust}$/L$_{FUV}$ ratios of z=0 and z=0.6 galaxies,
suggesting no evolution for the internal extinction in
UV galaxies of the same luminosity. Indeed,
for both samples, the mean L$_{dust}$/L$_{FUV}$ ratios
do not show significant dependence on the UV luminosity.

\begin{figure}
\vbox {
  \begin{minipage}[l]{1.0\textwidth}
   \hbox{
      \begin{minipage}[l]{0.5\textwidth}
       {\centering \leavevmode \epsfxsize=\textwidth 
        \epsfbox{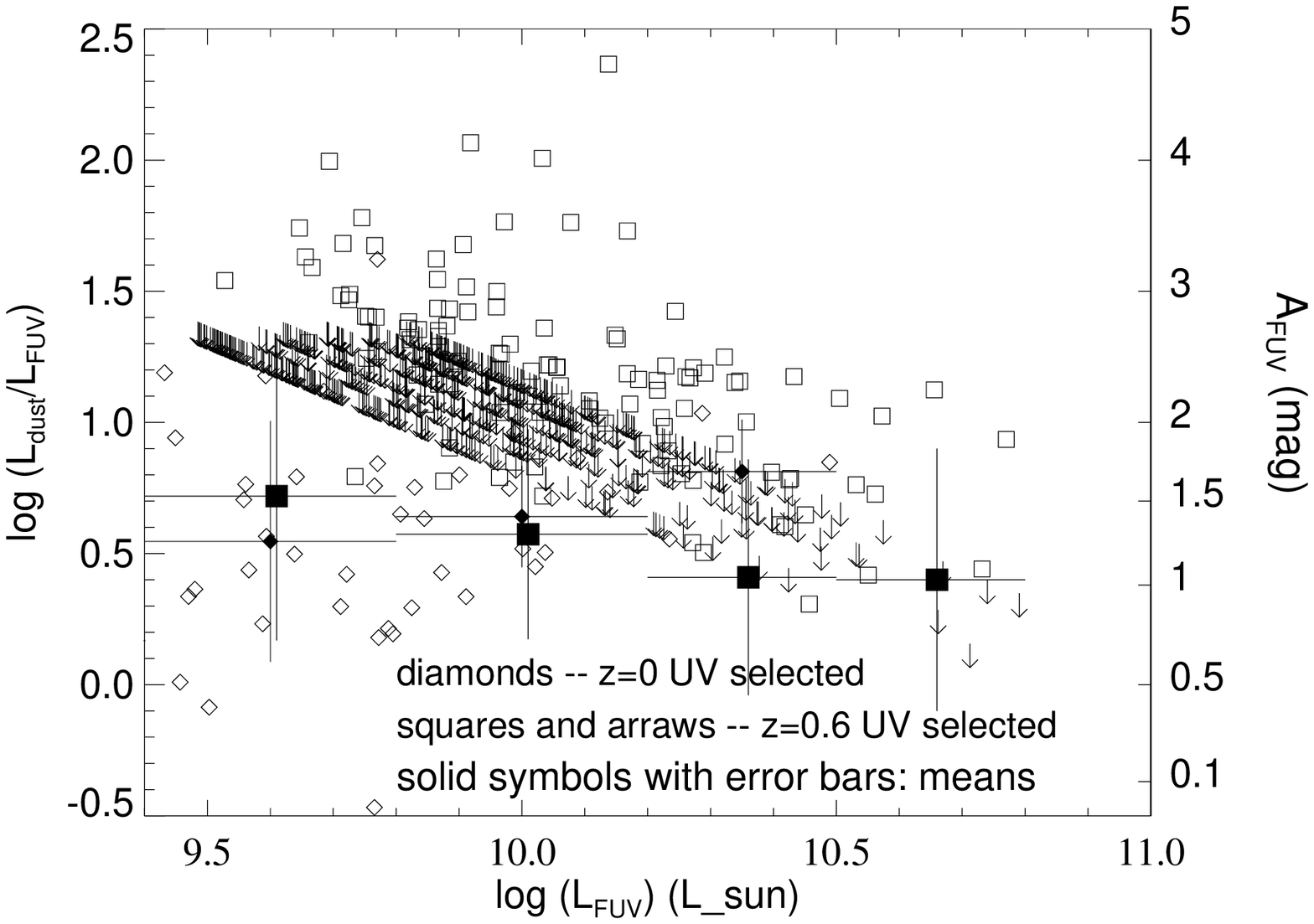}}
      \end{minipage} \  \hfill \
      \begin{minipage}[r]{0.5\textwidth}
       {\centering \leavevmode \epsfxsize=\textwidth 
        \epsfbox{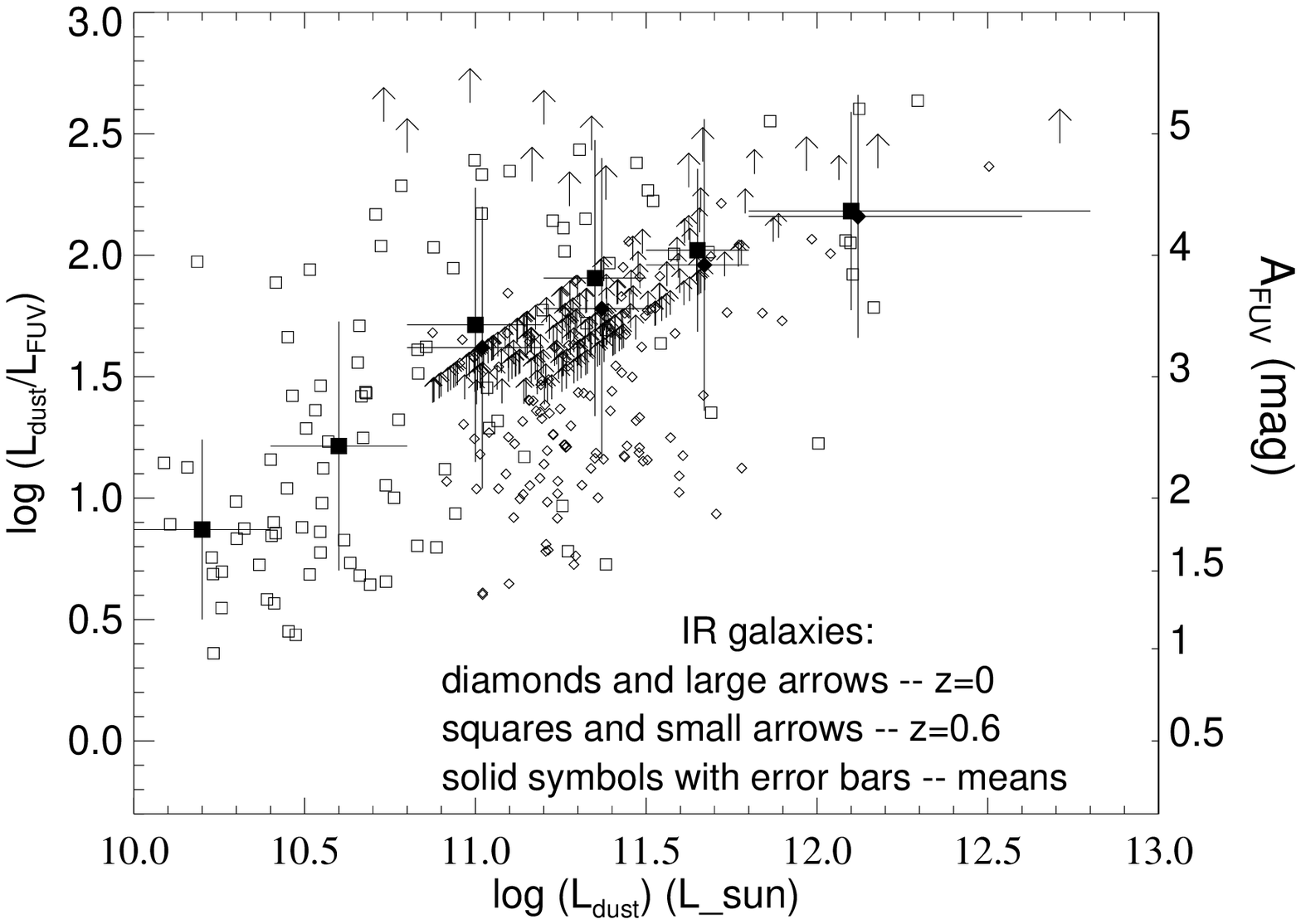}}
      \end{minipage} \  \hfill \
   }
  \end{minipage} \  \hfill \
\begin{minipage}[l]{1.0\textwidth}
\caption{L$_{dust}$/L$_{FUV}$ vs. luminosity plots of UV and IR galaxies:
comparisons between galaxies at z=0.6 and at z=0.}
\end{minipage}
}
\label{fig8}
\end{figure}

The NUV detection rate of  24$\mu m$ selected galaxies 
is 40\%. We use both stacking method and the survival
technique~\cite{feigelson85}~\cite{schmitt85}
to derive the means of L$_{dust}$/L$_{FUV}$ of these
galaxies in different L$_{dust}$ bins, the results are
plotted in Fig.8b. In contrast with
the UV galaxies,  both z=0.6 and z=0 IR galaxies
show strong dependence of the L$_{dust}$/L$_{FUV}$
with the luminosity. Indeed, IR galaxies in the luminosity
range covered by the $z=0.6$ sample have significantly
higher L$_{dust}$/L$_{FUV}$ ratios than those of
the UV galaxies (Fig.8a). On the other hand, as shown
by the mean ratios, the IR galaxies of z=0.6 have about the same
L$_{dust}$/L$_{FUV}$ as their z=0 counterparts of the same
IR luminosity.

The most serious uncertainty in the comparisons above is due to the
extrapolation from L$_{15\mu m}$ to L$_{dust}$.  The most direct way
to constrain this uncertainty is to look at the
real SEDs of the z=0.6 galaxies, in particular those
detected in Spitzer MIPS 70$\mu m$ (rest frame
43.75$\mu m$) and 160$\mu m$ (rest frame 100$\mu m$)
bands. In the sky region studied here, there is only
one z$\sim 0.6$ galaxy detected in 160$\mu m$ band.
This source has an SED very close to that of Arp220, which
has an L$_{dust}$/L$_{15\mu m}$ about 3 times higher of that
of Chary \& Elbaz value. Among the other 4 z$\sim 0.6$ 
galaxies detected in the 70$\mu m$ band, two have
SEDs similar to that of Mrk231 which has an 
 L$_{dust}$/L$_{15\mu m}$ about half of that
of Chary \& Elbaz value, other two have Arp220 type SEDs.
These results indicate that the uncertainty
due to variation of L$_{dust}$/L$_{15\mu m}$ is about
a factor of 2. 

\begin{figure}
\vbox {
  \begin{minipage}[l]{1.0\textwidth}
   \hbox{
      \begin{minipage}[l]{0.5\textwidth}
       {\centering \leavevmode \epsfxsize=\textwidth 
        \epsfbox{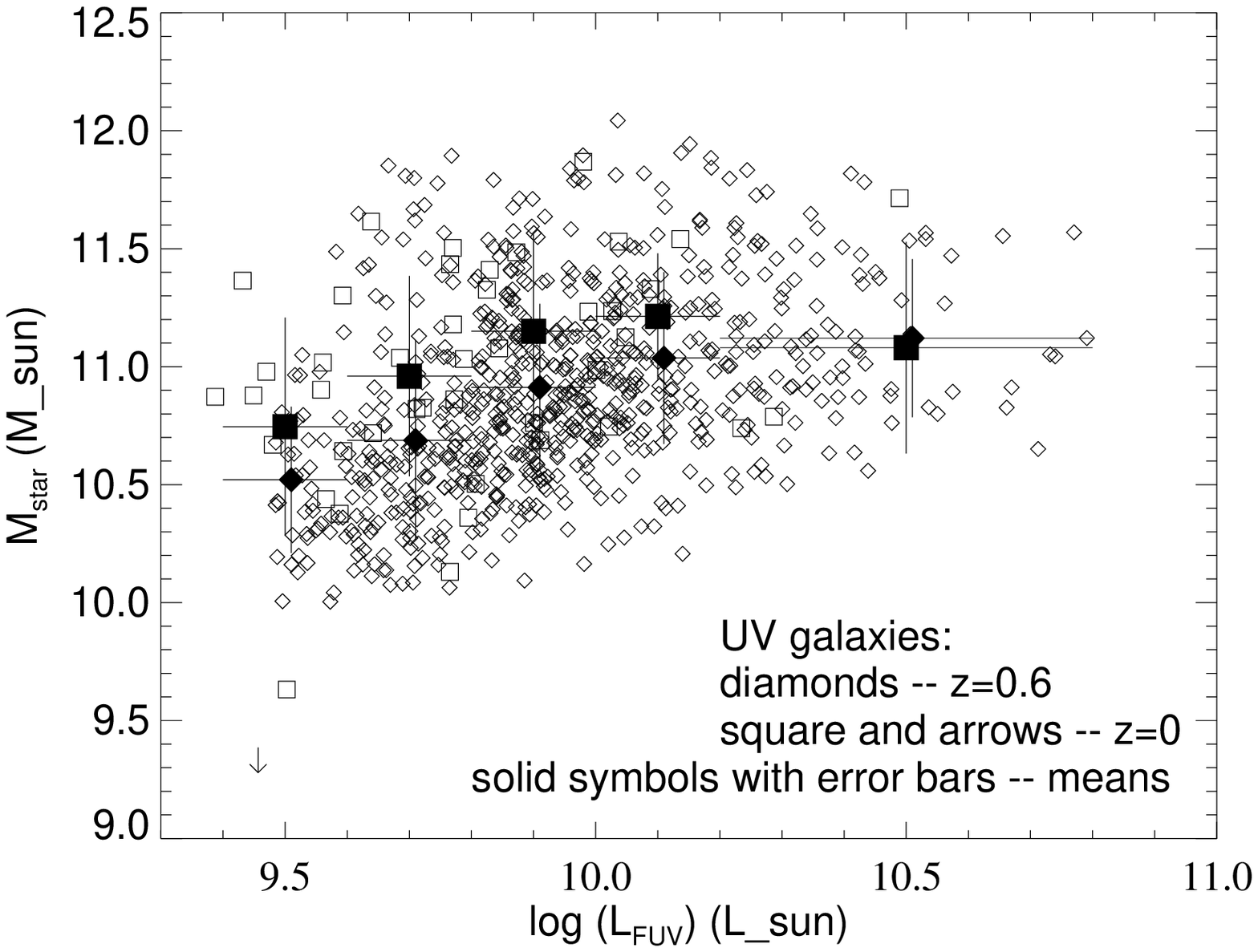}}
      \end{minipage} \  \hfill \
      \begin{minipage}[r]{0.5\textwidth}
       {\centering \leavevmode \epsfxsize=\textwidth 
        \epsfbox{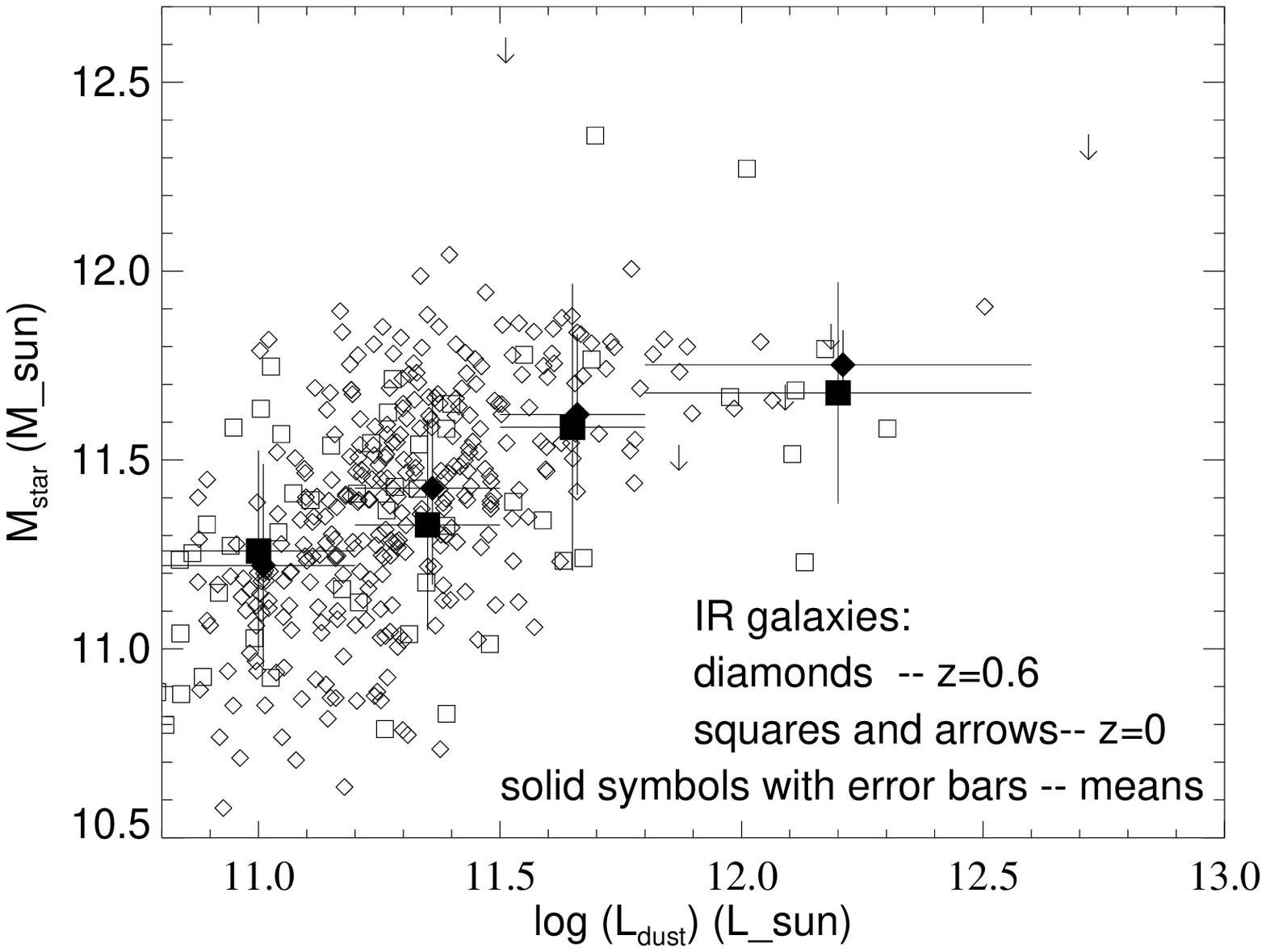}}
      \end{minipage} \  \hfill \
   }
  \end{minipage} \  \hfill \
\begin{minipage}[l]{1.0\textwidth}
\caption{Stellar mass vs. luminosity plots of UV and IR galaxies:
comparisons between galaxies at z=0.6 and at z=0.}
\end{minipage}
}
\label{fig9}
\end{figure}

In Fig.9a we compare the stellar mass of z=0.6 UV galaxies of given UV
luminosity to their local counterparts. The stellar mass is estimated
using the rest frame K band luminosity. 
The solid diamonds are the mean of the z=0.6 galaxies, and the solid
squares are the means of the local UV galaxies. Except for the
last bin, in all other bins the stellar mass of z=0.6 UV galaxies is
about a factor of 2 less than that of the local UV galaxies. This is
equivalent to a 2 times higher specific star formation rate in a given
FUV luminosity. The comparison of the stellar mass of
z=0.6 and z=0 IR selected galaxies is plotted in Fig.9b. 
Different from UV galaxies, there is no
evidence for any evolution in the stellar mass of IR selected
galaxies.

\section{Summary}
By selection, UV and IR galaxies have very
different characteristic IR/UV ratios. 
The morphological and stellar mass distributions of
UV and IR galaxies have good overlaps.
IR galaxies brighter than $L_{60\mu m}=10^{11}$ L$_\sun$
are severely under-represented in the UV sample, and
a population of low mass,
low luminosity UV galaxies are largely missing in the IR sample.
In the local universe, the contribution from bright IR galaxies 
(LIRGs and ULIRGs) and from the `IR-quiet' UV galaxies to the 
total star formation are negligible, the selection effect
in the UV and IR samples does not introduce significant bias.

Star forming galaxies at intermediate redshifts (z$\sim 0.6$)
do not show significant evolution in the dust attenuation
compared to their z=0 counterparts of same luminosity. The 
strong evolution for the dust attenuation derived from
the ratios of mean UV and IR luminosity density at different
redshifts~\cite{buat04} is therefore due to the strong
luminosity evolution of star forming galaxies and the
dependence of the dust attenuation on $L_{tot}$.
There is evidence for decrease of stellar mass of
UV galaxies with redshift, indicating a continuous
assembly of these galaxies in the recent history of the universe.
No such evidence for the IR galaxies.

\section*{Acknowledgments}
Collaborations with Michael Rowan-Robinson, Jorge Iglesias-P\'{a}ramo,
Tsutomu T. Takeuchi, and other members of GALEX team and SWIRE team
are acknowledged.

\end{document}